\documentclass[aoas,seceqn,nameyear,dvips]{arximspdf}
\usepackage{graphicx}

\doi{10.1214/10-AOAS388}
\volume{5}
\issue{1}
\pubyear{2011}
\firstpage{232}
\lastpage{253}

\makeatletter
\newcommand{\x}{\mathbf{x}}
\newcommand{\X}{\mathbf{X}}
\newcommand{\y}{\mathbf{y}}

\newcommand{\W}{\mathbf{W}}
\newcommand{\I}{\mathbf{I}}
\newcommand{\D}{\mathbf{D}}
\renewcommand{\r}{\mathbf{r}}
\renewcommand{\u}{\mathbf{u}}

\newcommand{\bb}{\bolds{\beta}}
\newcommand{\be}{\bolds{\eta}}
\newcommand{\bp}{\bolds{\pi}}

\newcommand{\bh}{\hat{\beta}}
\newcommand{\bbh}{\hat{\bolds{\beta}}}

\newcommand{\rj}{\mathbf{r}_{-j}}
\newcommand{\bbj}{\bolds{\beta}_{-j}}
\newcommand{\Xj}{\mathbf{X}_{-j}}
\newcommand{\yy}{\tilde{\mathbf{y}}}

\newcommand{\E}{\mathrm{E}}

\newcommand{\abs}[1]{\vert#1\vert}

\newtheorem{lemmamas}{Lemma}
\newtheorem{prop}{Proposition}
\newcommand{\eqref}[1]{(\ref{#1})}
\makeatother

\begin{document}
\begin{frontmatter}

\title{Coordinate descent algorithms for nonconvex penalized regression, with applications to biological
feature selection}
\runtitle{Coordinate descent for nonconvex penalized regression}

\begin{aug}
\author[A]{\fnms{Patrick} \snm{Breheny}\corref{}\thanksref{t1}\ead[label=e1]{patrick.breheny@uky.edu}}
\and
\author[B]{\fnms{Jian} \snm{Huang}\thanksref{t2}\ead[label=e2]{jian-huang@uiowa.edu}}

\thankstext{t1}{Supported in part by NIH Grant T32GM077973-05.}
\thankstext{t2}{Supported in part by NIH Grant R01CA120988 and NSF Grant DMS-08-05670.}
\runauthor{P. Breheny and J. Huang}
\affiliation{University of Kentucky and University of Iowa}
\address[A]{Department of Biostatistics\\
Department of Statistics\\
University of Kentucky\\
121 Washington Ave., Room 203C\\
Lexington, Kentucky 40536-0003\\USA\\
\printead{e1}} %adresu isvedimo komanda gale!

\address[B]{Department of Statistics\\
and Actuarial Sciences\\
Department of Biostatistics\\
University of Iowa\\
241 Schaeffer Hall\\
Iowa City, Iowa 52242\\USA\\
\printead{e2}}
\end{aug}

% HISTORY:
\received{\smonth{11} \syear{2009}}
\revised{\smonth{7} \syear{2010}}

% ABSTRACT
\begin{abstract}
A number of variable selection methods have been proposed involving
nonconvex penalty functions.  These methods, which include the smoothly
clipped absolute deviation (SCAD) penalty and the minimax concave
penalty (MCP), have been demonstrated to have attractive theoretical
properties, but model fitting is not a straightforward task, and the
resulting solutions may be unstable.  Here, we demonstrate the
potential of coordinate descent algorithms for fitting these models,
establishing theoretical convergence properties and demonstrating that
they are significantly faster than competing approaches.  In addition,
we demonstrate the utility of convexity diagnostics to determine
regions of the parameter space in which the objective function is
locally convex, even though the penalty is not.  Our simulation study
and data examples indicate that nonconvex penalties like MCP and SCAD
are worthwhile alternatives to the lasso in many applications.  In
particular, our numerical results suggest that MCP is the preferred
approach among the three methods.
\end{abstract}

% KEYWORDS
\begin{keyword}
\kwd{Coordinate descent}
\kwd{penalized regression}
\kwd{lasso}
\kwd{SCAD}
\kwd{MCP}
\kwd{optimization}.
\end{keyword}

\end{frontmatter}

%s1 ###
\section{Introduction}

Variable selection is an important issue in regression.  Typically,
measurements are obtained for a large number of potential predictors in
order to avoid missing an important link between a predictive factor
and the outcome.  This practice has only increased in recent years, as
the low cost and easy implementation of automated methods for data
collection and storage has led to an abundance of problems for which
the number of variables is large in comparison to the sample size.

To reduce variability and obtain a more interpretable model, we often
seek a smaller subset of important variables.  However, searching
through subsets of potential predictors for an adequate smaller model
can be unstable [\citet{Breiman1996}] and is computationally unfeasible
even in modest dimensions.

To avoid these drawbacks, a number of penalized regression methods have
been proposed in recent years that perform subset selection in a
continuous fashion.  Penalized regression procedures accomplish this by
shrinking coefficients toward zero in addition to setting some
coefficients exactly equal to zero (thereby selecting the remaining
variables).  The most popular penalized regression method is the lasso
[\citet{Tibshirani1996}].  Although the lasso has many attractive
properties, the shrinkage introduced by the lasso results in
significant bias toward 0 for large regression coefficients.

Other authors have proposed alternative penalties, designed to diminish
this bias.  Two such proposals are the smoothly clipped absolute
deviation (SCAD) penalty [\citet{Fan2001}] and the mimimax concave
penalty [MCP; \citet{Zhang2010}].  In proposing SCAD and MCP, their
authors established that SCAD and MCP regression models have the
so-called \textit{oracle property}, meaning that, in the asymptotic
sense, they perform as well as if the analyst had known in advance
which coefficients were zero and which were nonzero.

However, the penalty functions for SCAD and MCP are nonconvex, which
introduces numerical challenges in fitting these models.  For the
lasso, which does possess a convex penalty, least angle regression
[LARS; \citet{Efron2004a}] is a remarkably efficient method for
computing an entire path of lasso solutions in the same order of time
as a least squares fit.  For nonconvex penalties, \citet{Zou2008} have
proposed making a local linear approximation (LLA) to the penalty,
thereby yielding an objective function that can be optimized using the
LARS algorithm.

More recently, coordinate descent algorithms for fitting
lasso-penalized models have been shown to be competitive with the LARS
algorithm, particularly in high dimensions
[\citet{Friedman2007}; \citet{Wu2008}; \citet{Friedman2010}].  In this paper we
investigate the application of coordinate descent algorithms to SCAD
and MCP regression models, for which the penalty is nonconvex.  We
provide implementations of these algorithms through the publicly
available \texttt{R} package, \texttt{ncvreg} (available at
\url{http://cran.r-project.org}).

Methods for high-dimensional regression and variable selection have
applications in many scientific fields, particularly those in
high-throughput biomedical studies.  In this article we apply the
methods to two such studies---a genetic association study and a gene
expression study---each possessing a different motivation for sparse
regression models.  In genetic association studies, very few genetic
markers are expected to be associated with the phenotype of interest.
Thus, the underlying data-generating process is likely to be highly
sparse, and sparse regression methods likely to perform well.  Gene
expression studies may also have sparse underlying representations;
however, even when they are relatively ``dense,'' there may be separate
motivations for sparse regression models.  For example, data from
microarray experiments may be used to discover biomarkers and design
diagnostic assays.  To be practical in clinical settings, such assays
must use only a small number of probes [\citet{Yu2007a}].

In Section \ref{section:linear} we describe algorithms for fitting linear regression
models penalized by MCP and SCAD, and discuss their convergence.  In
Section \ref{sec:logistic} we discuss the modification of those algorithms for fitting
logistic regression models.  Issues of stability, local convexity and
diagnostic measures for investigating these concerns are discussed,
along with the selection of tuning parameters, in Section \ref{section:convexity}.  The
numerical efficiency of the proposed algorithm is investigated in
Section \ref{section:sim} and compared with LLA algorithms.  The statistical properties
of lasso, SCAD and MCP are also investigated and compared using
simulated data (Section \ref{section:sim}) and applied to biomedical data (Section~\ref{section:app}).

%s2 ###
\section{Linear regression with nonconvex penalties}\label{section:linear}

Suppose we have $n$ observations, each of which contains measurements
of an outcome $y_i$ and $p$ features $\{x_{i1},\ldots,x_{ip}\}$.  We
assume without loss of generality that the features have been
standardized such that $\sum_{i=1}^n x_{ij}=0$, $n^{-1} \sum_{i=1}^n
x_{ij}^2 =1$ and $\sum_{i=1}^n y_i = 0$.  This ensures that the
penalty is applied equally to all covariates in an equivariant manner,
and eliminates the need for an intercept.  This is standard practice in
regularized estimation; estimates are then transformed back to their
original scale after the penalized models have been fit, at which point
an intercept is introduced.

We will consider models in which the expected value of the outcome
depends on the covariates through the linear function $\E(\y) = \be =
\X\bb$.  The problem of interest involves estimating the vector of
regression coefficients $\bb$.  Penalized regression methods accomplish
this by minimizing an objective function $Q$ that is composed of a loss
function plus a penalty function.  In this section we take the loss
function to be squared error loss:
%e2.1 ###
\begin{equation}\label{eq:q-lin}
Q_{\lambda,\gamma}(\bb)=\frac{1}{2n}\sum_{i=1}^n(y_i-\eta_i)^2+\sum_{j=1}^p p_{\lambda,\gamma}(\abs{\beta_j}),
\end{equation}
where $p_{\lambda,\gamma}(\cdot)$ is a function of the coefficients
indexed by a parameter $\lambda$ that controls the tradeoff between the
loss function and penalty, and that also may be shaped by one or more
tuning parameters $\gamma$.  This approach produces a spectrum of
solutions depending on the value of $\lambda$; such methods are often
referred to as regularization methods, with $\lambda$ the
regularization parameter.

To find the value of $\bb$ that optimizes \eqref{eq:q-lin}, the LLA
algorithm makes a linear approximation to the penalty, then uses the
LARS algorithm to compute the solution.  This process is repeated
iteratively\footnote{In \citet{Zou2008}, the authors also discuss a
one-step version of LLA starting from an initial estimate satisfying
certain conditions.  Although this is an interesting topic, we focus
here on algorithms that minimize the specified MCP/SCAD objective
functions and thereby possess the oracle properties demonstrated in
\citet{Fan2001} and \citet{Zhang2010} without requiring a separate
initial estimator.} until convergence for each value of $\lambda$ over
a grid.  Details of the algorithm and its implementation may be found
in \citet{Zou2008}.

The LLA algorithm is inherently inefficient to some extent, in that it
uses the path-tracing LARS algorithm to produce updates to the
regression coefficients.  For example, over a grid of 100 values for
$\lambda$ that averages 10 iterations until convergence at each point,
the LLA algorithm must calculate 1000 lasso paths to produce a single
approximation to the MCP or SCAD path.

An alternative to LLA is to use a coordinate descent approach.
Coordinate descent algorithms optimize a target function with respect
to a single parameter at a time, iteratively cycling through all
parameters until convergence is reached.  The idea is simple but
efficient---each pass over the parameters requires only $O(np)$
operations.  If the number of iterations is smaller than $p$, the
solution is reached with even less computation burden than the $np^2$
operations required to solve a linear regression problem by QR
decomposition.  Furthermore, since the computational burden increases
only linearly with $p$, coordinate descent algorithms can be applied to
very high-dimensional problems.

Coordinate descent algorithms are ideal for problems that have a simple
closed form solution in a single dimension but lack one in higher
dimensions.  The basic structure of a coordinate descent algorithm is,
simply: For $j$ in $\{1,\ldots,p\}$, to partially optimize $Q$ with
respect to $\beta_j$ with the remaining elements of $\bb$ fixed at
their most recently updated values.  Like the LLA algorithm, coordinate
descent algorithms iterate until convergence is reached, and this
process is repeated over a grid of values for $\lambda$ to produce a
path of solutions.

The efficiency of coordinate descent algorithms comes from two sources:
(1)~updates can be computed very rapidly, and (2) if we are computing a
continuous path of solutions (see Section~\ref{section:path}), our
initial values will never be far from the solution and few iterations
will be required.  Rapid updates are possible because the minimization
of $Q$ with respect to $\beta_j$ can be obtained from the univariate
regression of the current residuals $\r = \y-\X\bb$ on $\x_j$, at a
cost of $O(n)$ operations.  The specific form of these updates depends
on the penalty and whether linear or logistic regression is being
performed, and will be discussed further in their respective sections.

In this section we describe coordinate descent algorithms for least
squares regression penalized by SCAD and MCP, as well as investigate
the convergence of these algorithms.

%s2.1 ###
\subsection{MCP}

\begin{figure}

\includegraphics{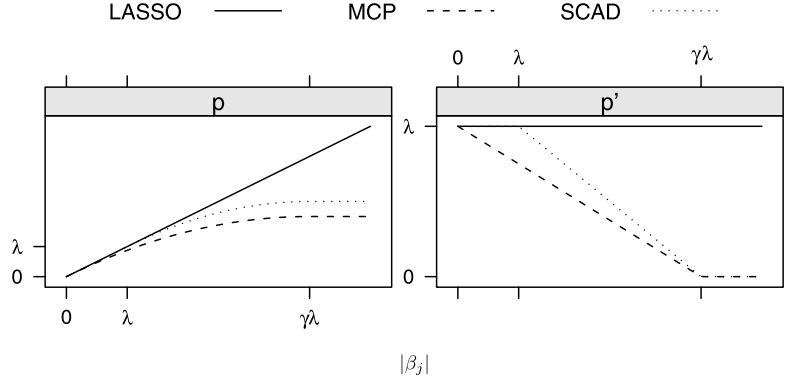}

\caption{Shapes of the lasso, SCAD and MCP penalty
functions. The panel on the left plots the penalties themselves,
whereas the panel on the right plots the derivative of the penalty.
Note that none of the penalties are differentiable at $\beta_j=0$.}\label{fig:shapes}
\end{figure}

Zhang (\citeyear{Zhang2010}) proposed the MCP, defined on $[0,\infty)$ by
\begin{eqnarray}\label{MCP}
p_{\lambda,\gamma}(\theta)&=&\cases{
\displaystyle\lambda\theta-\frac{\theta^2}{2\gamma},&\quad if $\theta \leq \gamma\lambda$,\cr
\displaystyle\frac{1}{2} \gamma\lambda^2,&\quad if $\theta
>\gamma\lambda$,}\nonumber
\\[-8pt]\\[-8pt]
p_{\lambda,\gamma}'(\theta)&=&\cases{
\displaystyle\lambda - \frac{\theta}{\gamma},&\quad if $\theta \leq \gamma\lambda$,\cr
0,&\quad if $\theta > \gamma\lambda$}\nonumber
\end{eqnarray}
for $\lambda \geq 0$ and $\gamma>1$.  The rationale behind the penalty
can be understood by considering its derivative: MCP begins by applying
the same rate of penalization as the lasso, but continuously relaxes
that penalization until, when $\theta > \gamma\lambda$, the rate of
penalization drops to 0.  The penalty is illustrated in
Figure~\ref{fig:shapes}.

%f1 ###

The rationale behind the MCP can also be understood by considering its
univariate solution.  Consider the simple linear regression of $\y$
upon $\x$, with unpenalized least squares solution $z=n^{-1}\x'\y$
(recall that $\x$ has been standardized so that $\x'\x=n$).  For this
simple linear regression problem, the MCP estimator has the following
closed form:
%e2.2 ###
\begin{equation}\label{eq:mcp-lin-sol}
\bh = f_\mathrm{MCP}(z,\lambda,\gamma) =\cases{
\displaystyle\frac{S(z,\lambda)}{1-1/\gamma}, &\quad if $\abs{z} \leq \gamma\lambda$,\cr
z,&\quad if $\abs{z} > \gamma\lambda$,}
\end{equation}
where $S$ is the soft-thresholding operator [\citet{Donoho1994}] defined
for $\lambda \geq 0$ by
%e2.3 ###
\begin{equation}
\label{soft}
S(z,\lambda) = \cases{
z-\lambda,&\quad if $z > \lambda$, \cr
0,&\quad if $\abs{z} \leq \lambda$, \cr
z+\lambda,&\quad if $z < -\lambda$.}
\end{equation}

Noting that $S(z,\lambda)$ is the univariate solution to the lasso, we
can observe by comparison that MCP scales the lasso solution back
toward the unpenalized solution by an amount that depends on $\gamma$.
As $\gamma \to \infty$, the MCP and lasso solutions are the same.  As
$\gamma \to 1$, the MCP solution becomes the hard thresholding estimate
$zI_{\abs{z}>\lambda}$.  Thus, in the univariate sense, the MCP
produces the ``firm shrinkage'' estimator of \citet{Gao1997}.

In the special case of an orthonormal design matrix, subset selection
is equivalent to hard-thresholding, the lasso is equivalent to
soft-thresholding, and MCP is equivalent to firm-thresholding.  Thus,
the lasso may be thought of as performing a kind of multivariate
soft-thresholding, subset selection as multivariate hard-thresholding,
and the MCP as multivariate firm-threshol\-ding.

The univariate solution of the MCP is employed by the coordinate
descent algorithm to obtain the coordinate-wise minimizer of the
objective function.  In this setting, however, the role of the
unpenalized solution is now played by the unpenalized regression of
$\x_j$'s partial residuals on $\x_j$, and denoted $z_j$.  Introducing
the notation $-j$ to refer to the portion that remains after the
$j$th column or element is removed, the partial residuals of $\x_j$
are $\rj=\y-\Xj\bbj$, where $\bbj$ is the most recently updated value
of $\bb$.  Thus, at step $j$ of iteration $m$, the following three
calculations are made:

\begin{itemize}
\item[(1)] calculate
\[
z_j=n^{-1}\x_j'\rj= n^{-1}\x_j'\r+\beta_j^{(m)},
\]
\item[(2)] update $\beta_j^{(m+1)}\gets
f_\mathrm{MCP}(z_j,\lambda,\gamma)$,
\item[(3)] update $\r\gets\r-(\beta_j^{(m+1)}-\beta_j^{(m)})\x_j$,
\end{itemize}
where the last step ensures that $\r$ always holds the current values
of the residuals.  Note that $z_j$ can be obtained by regressing $\x_j$
on either the partial residuals or the current residuals; using the
current residuals is more computationally efficient, however, as it
does not require recalculating partial residuals for each update.

%s2.2 ###
\subsection{SCAD}\label{section:scad}

The SCAD penalty \citet{Fan2001} defined on $[0,\infty)$ is given by
\begin{eqnarray}\label{SCAD}
p_{\lambda, \gamma}(\theta)
&=&
\cases{
\lambda \theta, &\quad if $\theta \leq \lambda$,\cr
\displaystyle\frac{\gamma\lambda\theta-0.5(\theta^2+\lambda^2)}{\gamma-1}, &\quad if $\lambda < \theta \leq \gamma\lambda$,\cr
\displaystyle\frac{\lambda^2(\gamma^2-1)}{2(\gamma-1)}, &\quad if $\theta >\gamma\lambda$,}\nonumber
\\[-8pt]\\[-8pt]
p_{\lambda}'(\theta)
&=&
\cases{
\lambda, &\quad if $\theta \leq \lambda$,\cr
\displaystyle\frac{\gamma\lambda-\theta}{\gamma-1}, &\quad if $\lambda < \theta \leq\gamma\lambda$,\cr
0, &\quad if $\theta > \gamma\lambda$}\nonumber
\end{eqnarray}
for $\lambda \geq 0$ and $\gamma>2$.  The rationale behind the penalty
is similar to that of MCP.  Both penalties begin by applying the same
rate of penalization as the lasso, and reduce that rate to 0 as
$\theta$ gets further away from zero; the difference is in the way that
they make the transition.  The SCAD penalty is also illustrated in
Figure~\ref{fig:shapes}.

The univariate solution for a SCAD-penalized regression coefficient is
as follows, where again $z$ is the unpenalized linear regression
solution:
%e2.4 ###
\begin{equation}\label{eq:scad-lin-sol}
\bh=f_\mathrm{SCAD}(z,\lambda,\gamma)=\cases{
S(z,\lambda),&\quad if $\abs{z}\leq 2\lambda$,\cr
\displaystyle\frac{S(z,\gamma\lambda/({\gamma-1}))}{1-1/(\gamma-1)},&\quad if $2\lambda<\abs{z}\leq\gamma\lambda$,\cr
z,&\quad if $\abs{z}>\gamma\lambda$.}
\end{equation}
This solution is similar to, although not the same as, the MCP
solution/firm shrinkage estimator.  Both penalties rescale the
soft-thresholding solution toward the unpenalized solution.  However,
although SCAD has soft-thresholding as the limiting case when $\gamma
\to \infty$, it does not have hard thresholding as the limiting case
when $\gamma \to 2$.  This univariate solution plays the same role in
the coordinate descent algorithm for SCAD as
$f_\mathrm{MCP}(z_j,\lambda,\gamma)$ played for MCP.

%s2.3 ###
\subsection{Convergence}
\label{section:convergence}

We now consider the convergence of coordinate descent algorithms for
SCAD and MCP.  We begin with the following lemma.

\begin{lemmamas}\label{1lema}
Let $Q_{j,\lambda,\gamma}(\beta)$ denote the objective function $Q$,
defined in \eqref{eq:q-lin}, as a function of the single variable
$\beta_j$, with the remaining elements of $\bb$ fixed.  For the SCAD
penalty with $\gamma > 2$ and for the MCP with $\gamma > 1$,
$Q_{j,\lambda,\gamma}(\beta)$ is a convex function of $\beta_j$ for all
$j$.
\end{lemmamas}

From this lemma, we can establish the following convergence properties
of coordinate descent algorithms for SCAD and MCP.

\begin{prop}\label{prop:cd}
Let $\{\bb^{(k)}\}$ denote the sequence of coefficients
produced at each iteration of the coordinate descent algorithms for
SCAD and MCP.  For all $k=0,1,2,\ldots,$
\[
Q_{\lambda,\gamma}\bigl(\bb^{(k+1)}\bigr) \leq Q_{\lambda,\gamma}\bigl(\bb^{(k)}\bigr).
\]
Furthermore, the sequence is guaranteed to converge to a point that is
both a local minimum and a global coordinate-wise minimum of
$Q_{\lambda,\gamma}$.
\end{prop}

Because the penalty functions of SCAD and MCP are not convex, neither
the coordinate descent algorithms proposed here nor the LLA algorithm
are guaranteed to converge to a global minimum in general.  However, it
is possible for the objective function $Q$ to be convex with respect to
$\bb$ even though it contains a nonconvex penalty component.  In
particular, letting $c_*$ denote the minimum eigenvalue of
$n^{-1}\X'\X$, the MCP objective function is convex if $\gamma >
1/c_*$, while the SCAD objective function is convex if $\gamma > 1 +
1/c_*$ [\citet{Zhang2010}].  If this is the case, then the coordinate
descent algorithms converge to the global minimum.
Section~\ref{section:convexity} discusses this issue further with
respect to high-dimensional settings.

%s2.4 ###
\subsection{Pathwise optimization and initial values}
\label{section:path}

Usually, we are interested in obtaining $\bbh$ not just for a single
value of $\lambda$, but for a range of values extending from a maximum
value $\lambda_{\max}$ for which all penalized coefficients are 0
down to $\lambda=0$ or to a minimum value $\lambda_{\min}$ at which
the model becomes excessively large or ceases to be identifiable.  When
the objective function is strictly convex, the estimated coefficients
vary continuously with $\lambda \in [\lambda_{\min}, \
\lambda_{\max}]$ and produce a path of solutions regularized by
$\lambda$.  Examples of such paths may be seen in Figures~\ref{fig:dgn}
and \ref{fig:app-gamma}.

Because the coefficient paths are continuous (under strictly convex
objective functions), a reasonable approach to choosing initial values
is to start at one extreme of the path and use the estimate $\bbh$ from
the previous value of $\lambda$ as the initial value for the next value
of $\lambda$.  For MCP and SCAD, we can easily determine
$\lambda_{\max}$, the smallest value for which all penalized
coefficients are 0.  From \eqref{eq:mcp-lin-sol} and
\eqref{eq:scad-lin-sol}, it is clear that $\lambda_{\max} =
z_{\max}$, where $z_{\max}=\max_j\{n^{-1}\x_j'\y\}$ [for
logistic regression (Section~\ref{sec:logistic}) $\lambda_{\max}$
is again equal to ${\max}_j\{z_j\}$, albeit with $z_j$ as defined
in \eqref{eq:logistic-z} and the quadratic approximation taken with
respect to the intercept-only model].  Thus, by starting at
$\lambda_{\max}$ with $\bb^{(0)}=\mathbf{0}$ and proceeding toward
$\lambda_{\min}$, we can ensure that the initial values will never
be far from the solution.

For all the numerical results in this paper, we follow the approach of
\citet{Friedman2010} and compute solutions along a grid of 100
$\lambda$ values that are equally spaced on the log scale.

%s3 ###
\section{Logistic regression with nonconvex penalties}
\label{sec:logistic}

For logistic regression, it is not possible to eliminate the need for
an intercept by centering the response variable.  For logistic
regression, then, $\y$ will denote the original vector of 0--1
responses.  Correspondingly, although $\X$ is still standardized, it
now contains an unpenalized column of 1's for the intercept, with
corresponding coefficient $\beta_0$.  The expected value of $\y$ once
again depends on the linear function $\be = \X\bb$, although the model
is now
%e3.1 ###
\begin{equation}
\label{eq:log-model} P(y_i=1|x_{i1},\ldots,x_{ip}) = \pi_i =
\frac{e^{\eta_i}}{1+e^{\eta_i}}.
\end{equation}
Estimation of the coefficients is now accomplished via minimization of
the objective function
%e3.2 ###
\begin{equation}
\label{eq:log-obj}
\qquad Q_{\lambda,\gamma}(\bb) = -\frac{1}{n}\sum_{i=1}^n
\{y_i \log{\pi_i} + (1-y_i)\log{(1-\pi_i)}\} + \sum_{j=1}^p
p_{\lambda,\gamma}(\abs{\beta_j}).
\end{equation}

Minimization can be approached by first obtaining a quadratic
approximation to the loss function based on a Taylor series expansion
about the value of the regression coefficients at the beginning of the
current iteration, $\bb^{(m)}$.  Doing so results in the familiar form
of the iteratively reweighted least squares algorithm commonly used to
fit generalized linear models [\citet{McCullagh1989}]:
%e3.3 ###
\begin{equation}
\label{eq:log-obj-quad} Q_{\lambda,\gamma}(\bb) \approx
\frac{1}{2n}(\yy - \X\bb)'\W(\yy-\X\bb) + \sum_{j=1}^p
p_{\lambda,\gamma}(\abs{\beta_j}),
\end{equation}
where $\yy$, the working response, is defined by
\[
\yy = \X\bb^{(m)} + \W^{-1}(\y-\bp)
\]
and $\W$ is a diagonal matrix of weights, with elements
\[
w_i = \pi_i(1-\pi_i),
\]
and $\bp$ is evaluated at $\bb^{(m)}$.  We focus on logistic regression
here, but the same approach could be applied to fit penalized versions
of any generalized linear model, provided that the quantities $\yy$ and
$\W$ [as well as the residuals $\r \equiv \yy - \X\bb^{(m)}$, which
depend on $\yy$ implicitly] are replaced with expressions specific to
the particular response distribution and link function.

With this representation, the local linear approximation (LLA) and
coordinate descent (CD) algorithms can be extended to logistic
regression in a rather straightforward manner.  At iteration $m$, the
following two steps are taken:
\begin{longlist}
\item[(1)] Approximate the loss function based on $\bb^{(m)}$.
\item[(2)] Execute one iteration of the LLA/CD algorithm, obtaining
$\bb^{(m+1)}$.
\end{longlist}
These two steps are then iterated until convergence for each value of
$\lambda$.  Note that for the coordinate descent algorithm, step (2)
loops over all the covariates.  Note also that step 2 must now involve
the updating of the intercept term, which may be accomplished without
modification of the underlying framework by setting $\lambda=0$ for the
intercept term.

The local linear approximation is extended in a straightforward manner
to reweighted least squares by distributing $\W$, obtaining the
transformed covariates and response variable $\W^{1/2}\X$ and
$\W^{1/2}\yy$, respectively.  The implications for coordinate descent
algorithms are discussed in the next section.

Briefly, we note that, as is the case for the traditional iteratively
reweighted least squares algorithm applied to generalized linear
models, neither algorithm (LLA/CD) is guaranteed to converge for
logistic regression.  However, provided that adequate safeguards are in
place to protect against model saturation, we have not observed failure
to converge to be a problem for either algorithm.

%s3.1 ###
\subsection{Fixed scale solution}\label{s31}

The presence of observation weights changes the form of the
coordinate-wise updates.  Let $v_j = n^{-1}\x_j'\W\x_j$, and redefine
$\r = \W^{-1}(\y-\bp)$ and
%e3.5 ###
%e3.4 ###
\begin{eqnarray}
\label{eq:logistic-z}
z_j &=& \frac{1}{n} \x_j'\W(\yy - \Xj\bbj)\\
  &=& \frac{1}{n} \x_j'\W\r + v_j\beta_j^{(m)}.
\end{eqnarray}
Now, the coordinate-descent update for MCP is
%e3.6 ###
\begin{equation}
\label{mcp-log-sol}
\beta_j \gets \cases{
\displaystyle\frac{S(z_j,\lambda)}{v_j-1/\gamma},&\quad if $\abs{z_j} \leq v_j\gamma\lambda$,\cr
\displaystyle\frac{z_j}{v_j},&\quad if $\abs{z_j} > v_j\gamma\lambda$}
\end{equation}
 for $\gamma > 1/v_j$ and for SCAD,
%e3.7 ###
\begin{equation}
\label{scad-log-sol}
\beta_j \gets\cases{
\displaystyle\frac{S(z_j,\lambda)}{v_j},&\quad if $\abs{z_j} \leq \lambda(v_j+1)$,\cr
\displaystyle\frac{S(z_j,\gamma\lambda/(\gamma-1))}{v_j-1/(\gamma-1)},&\quad if $\lambda(v_j+1) < \abs{z_j} \leq v_j\gamma\lambda$, \cr
\displaystyle\frac{z_j}{v_j},&\quad if $\abs{z_j} > v_j\gamma\lambda$}
\end{equation}
for $\gamma > 1+1/v_j$.  Updating of $\r$ proceeds as in the linear
regression case.

As is evident from comparing
\eqref{eq:mcp-lin-sol}/\eqref{eq:scad-lin-sol} with
\eqref{mcp-log-sol}/\eqref{scad-log-sol}, portions of both numerator
and denominator are being reweighted in logistic regression.  In
comparison, for linear regression, $v_j$ is always equal to 1 and this
term drops out of the solution.

This reweighting, however, introduces some difficulties with respect to
the choice and interpretation of the $\gamma$ parameter.  In linear
regression, the scaling factor by which solutions are adjusted toward
their unpenalized solution is a constant [$1-1/\gamma$ for MCP,
$1-1/(\gamma-1)$ for SCAD] for all values of $\lambda$ and for each
covariate.  Furthermore, for standardized covariates, this constant has
a universal interpretation for all linear regression problems, meaning
that theoretical arguments and numerical simulations investigating
$\gamma$ do not need to be rescaled and reinterpreted in the context of
applied problems.

In logistic regression, however, this scaling factor is constantly
changing, and is different for each covariate.  This makes choosing an
appropriate value for $\gamma$ difficult in applied settings and robs
the parameter of a consistent interpretation.

To illustrate the consequences of this issue, consider an attempt to
perform logistic regression updates using $\gamma=3.7$, the value
suggested for linear regression in \citet{Fan2001}.  Because $w_i$
cannot exceed 0.25, $\gamma$ cannot exceed $1/v_j$ and the solution is
discontinuous and unstable.  Note that this problem does not arise from
the use of any particular algorithm---it is a direct consequence of
the poorly behaved objective function with this value of $\gamma$.

%s3.2 ###
\subsection{Adaptive rescaling}\label{s32}

To resolve these difficulties, we propose an adaptive rescaling of the
penalty parameter $\gamma$ to match the continually changing scale of
the covariates.  This can be accomplished by simply replacing
$p_{\lambda,\gamma}(\abs{\beta_j})$ with
$p_{\lambda,\gamma}(\abs{v_j\beta_j})$.  The algorithmic consequences
for the LLA algorithm are straightforward.  For coordinate descent, the
updating steps become simple extensions of the linear regression
solutions:
\[
\beta_j \gets \frac{f(z_j,\lambda,\gamma)}{v_j}.
\]

Note that, for MCP,
\[
\frac{S(z_j,\lambda)}{v_j(1-1/\gamma)} = \frac{S(z_j,\lambda)}{v_j -
1/\gamma^{*}},
\]
where $\gamma^{*} = \gamma / v_j$.  Thus, the adaptively rescaled solution
is still minimizing the objective function \eqref{eq:log-obj-quad},
albeit with an alternate set of shape parameters $\{\gamma_j^{*}\}$ that
are unknown until convergence is attained.

Note that rescaling by $v_j$ does not affect the magnitude of the
penalty ($\lambda$), only the range over which the penalty is applied
($\gamma$).  Is it logical to apply different scales to different
variables?  Keep in mind that, since $\x_j'\x_j=n$ for all $j$, the
rescaling factor $v_j$ will tend to be quite similar for all
covariates.  However, consider the case where a covariate is
predominantly associated with observations for which $\hat{\pi}_i$ is
close to 0 or 1.  For such a covariate, adaptive rescaling will extend
the range over which the penalization is applied.  This seems to be a
reasonable course of action, as large changes in this coefficient
produce only small changes in the model's fit, and provide less
compelling evidence of a covariate's importance.

SCAD does not have the property that its adaptively rescaled solution
is equal to a solution of the regular SCAD objective function with
different shape parameters.  This is due to the fact that the scale of
the penalty is tied to the scale of the coefficient by the $\theta <
\lambda$ clause.  One could make this clause more flexible by
reparameterizing SCAD so that $p'(\theta)=\lambda$ in the region
$\theta < \gamma_2\lambda$, where $\gamma_2$ would be an additional
tuning parameter.  In this generalized case, adaptively rescaled SCAD
would minimize a version of the original objective function in which
the $\gamma$ parameters are rescaled by $v_j$, as in the MCP case.

As we will see in Section \ref{section:sim}, this adaptive rescaling increases
interpretability and makes it easier to select $\gamma$.

%s4 ###
\section{Convexity, stability and selection of tuning parameters}
\label{section:convexity}

%s4.1 ###
\subsection{Convexity and stability}

As mentioned in Section~\ref{section:convergence}, the
MCP/SCAD-penalized linear regression objective function is convex
provided that $\gamma > 1/c_*$ (MCP) or $\gamma > 1 + 1/c_*$ (SCAD).
This result can be extended to logistic regression as well via the
following proposition.

\begin{prop}
\label{prop:logistic-convex} Let $c_*(\bb)$ denote the minimum
eigenvalue of $n^{-1}\X'\W\X$, where $\W$ is evaluated at $\bb$.  Then
the objective function defined in \eqref{eq:log-obj} is a convex
function of $\bb$ on the region where $c_*(\bb) > 1/\gamma$ for MCP,
and where $c_*(\bb) > 1/(\gamma-1)$ for SCAD.
\end{prop}

A convex objective function is desirable for (at least) two reasons.
First, for any algorithm (such as coordinate descent) which converges
to a critical point of the objective function, convexity ensures that
the algorithm converges to the unique global minimum.  Second,
convexity ensures that $\bbh$ is continuous with respect to $\lambda$,
which in turn ensures good initial values for the scheme described in
Section~\ref{section:path}, thereby reducing the number of iterations
required by the algorithm.

There are obvious practical benefits to using an algorithm that
converges rapidly to the unique global solution.  However, convexity
may also be desirable from a statistical perspective.  In the absence
of convexity, $\bbh$ is not necessarily continuous with respect to the
data---that is, a small change in the data may produce a large
change in the estimate.  Such estimators tend to have high variance
[\citet{Breiman1996}; \citet{Bruce1996}] in addition to being unattractive from
a logical perspective.  Furthermore, discontinuity with respect to
$\lambda$ increases the difficulty of choosing a good value for the
regularization parameter.

%s4.2 ###
\subsection{Local convexity diagnostics}

However, it is not always necessary to attain global convexity.  In
high-dimensional settings where $p > n$, global convexity is neither
possible nor relevant.  In such settings, sparse solutions for which
the number of nonzero coefficients is much lower than $p$ are desired.
Provided that the objective function is convex in the local region that
contains these sparse solutions, we will still have stable estimates
and smooth coefficient paths in the parameter space of interest.

Of considerable practical interest, then, is a diagnostic that would
indicate, for nonconvex penalties such as SCAD and MCP, which regions
of the coefficient paths are locally convex and which are not.  Here,
we introduce a definition for local convexity and a diagnostic measure
which can be easily computed from the data and which can indicate which
regions of the coefficient paths retain the benefits of convexity and
which do not.

Recall the conditions for global convexity: $\gamma$ must be greater
than $1/c_*$ for MCP ($1 + 1/c_*$ for SCAD), where $c_*$ denoted the
minimum eigenvalue of $n^{-1}\X'\X$.  We propose modifying this cutoff
so that only the covariates with nonzero coefficients (the covariates
which are ``active'' in the model) are included in the calculation of
$c_*$.  Note that neither $\gamma$ nor $\X$ change with $\lambda$.
What does vary with $\lambda$ is set of active covariates; generally
speaking, this will increase as $\lambda$ decreases (with
correlated/collinear data, however, exceptions are possible).  Thus,
local convexity of the objective function will not be an issue for
large $\lambda$, but may cease to hold as $\lambda$ is lowered past
some critical value $\lambda^*$.

Specifically, let $\bbh(\lambda)$ denote the minimizer of
\eqref{eq:q-lin} for a certain value of $\lambda$, $A(\lambda) =
\{j\dvtx\bh_j(\lambda) \ne 0\}$ denote the active set of covariates,
$U(\lambda)=A(\lambda) \cup A(\lambda^-)$ denote the set of covariates
that are either currently active given a value $\lambda$ or that will
become active imminently upon the lowering of $\lambda$ by an
infinitesimal amount, and let $\X_U(\lambda)$ denote the design matrix
formed from only those covariates for which $j \in U(\lambda)$, with
$c_*(\lambda)$ denoting the minimum eigenvalue of
$n^{-1}{\X_{U(\lambda)}}'\X_{U(\lambda)}$.  Now, letting
\[
\lambda^* =\inf\{\lambda\dvtx\gamma > 1/c_*(\lambda)\} \qquad \mbox{for
MCP}\hspace*{25.5pt}
\]
and
\[
\lambda^* =\inf\{\lambda\dvtx\gamma > 1+1/c_*(\lambda)\} \qquad \mbox{for SCAD,}
\]
we define the $\lambda$ interval over which the objective function is
``locally convex'' to be $(\infty,\lambda^*)$.  Correspondingly, the
objective function is locally nonconvex (and globally nonconvex) in the
region $[\lambda^*,0]$.  Because $c_*(\lambda)$ changes only when the
composition of $U(\lambda)$ changes, it is clear that $\lambda^*$ must
be a value of $\lambda$ for which $A(\lambda) \ne A(\lambda^-)$.

For logistic regression, let $c_*(\lambda)$ represent the minimum
eigenvalue of $n^{-1}{\X_{U(\lambda)}}'\W\X_{U(\lambda)} - \D_p$, where
$\W$ is evaluated at $\bbh(\lambda)$ and $\D_p$ is a diagonal matrix
that depends on the penalty function.  For the fixed scale estimation
of Section \ref{s31}, $\D_p$ has elements $\{1/\gamma\}$ for MCP and
$\{1/(\gamma-1)\}$ for SCAD, while for the adaptively rescaled
estimation of Section \ref{s32}, $\D_p$ has elements
$\{n^{-1}\x_k'\W\x_k/\gamma\}$ for MCP and
$\{n^{-1}\x_k'\W\x_k/(\gamma-1)\}$ for SCAD.  For $c_*(\lambda)$
defined in this manner, $\lambda^*$ equals the smallest value of
$\lambda$ such that $c_*(\lambda) > 0$.

%f2 ###
\begin{figure}

\includegraphics{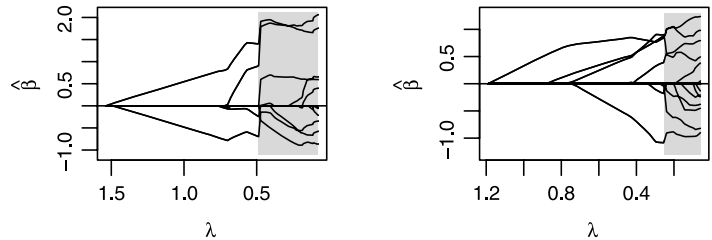}

\caption{Example MCP coefficient paths for simulated
data where $p>n$. The shaded region is the region in which the
objective function is not locally convex. Note that the solutions are
continuous and stable in the locally convex regions, but discontinuous
and erratic otherwise.}\label{fig:dgn}
\end{figure}

The practical benefit of these diagnostics can be seen in
Figure~\ref{fig:dgn}, which depicts examples of coefficient paths from
simulated data in which $n=20$ and $p=50$.  As is readily apparent,
solutions are smooth and well behaved in the unshaded, locally convex
region, but discontinuous and noisy in the shaded region which lies to
the right of $\lambda^*$.  Figure~\ref{fig:dgn} contains only MCP
paths; the corresponding figures for SCAD paths look very similar.
Figure~\ref{fig:dgn} displays linear regression paths; paths for
logistic regression can be seen in Figure~\ref{fig:app-gamma}.

The noisy solutions in the shaded region of Figure~\ref{fig:dgn} may
arise from numerical convergence to suboptimal solutions, inherent
statistical variability arising from minimization of a nonconvex
function, or a combination of both; either way, however, the figure
makes a compelling argument that practitioners of regression methods
involving nonconvex penalties should know which region their solution
resides in.

%s4.3 ###
\subsection{Selection of $\gamma$ and $\lambda$}
\label{section:selection}

Estimation using MCP and SCAD models depends on the choice of the
tuning parameters $\gamma$ and $\lambda$.  This is usually accomplished
with cross-validation or using an information criterion such as AIC or
BIC.  Each approach has its shortcomings.

Information criteria derived using asymptotic arguments for unpenalized
regression models are on questionable ground when applied to penalized
regression problems where $p > n$.  Furthermore, defining the number of
parameters in models with penalization and shrinkage present is
complicated and affects lasso, MCP and SCAD differently.  Finally, we
have observed that AIC and BIC have a tendency, in some settings, to
select local mimima in the nonconvex region of the objective function.

Cross-validation does not suffer from these issues; on the other hand,
it is computationally intensive, particularly when performed over a
two-dimensional grid of values for $\gamma$ and $\lambda$, some of
which may not possess convex objective functions and, as a result,
converge slowly.  This places a barrier in the way of examining the
choice of $\gamma$, and may lead practitioners to use default values
that are not appropriate in the context of a particular problem.

It is desirable to select a value of $\gamma$ that produces
parsimonious models while avoiding the aforementioned pitfalls of
nonconvexity.  Thus, we suggest the following hybrid approach, using
BIC, cross-validation and convexity diagnostics in combination, which
we have observed to work well in practice.  For a path of solutions
with a given value of $\gamma$, use AIC/BIC to select $\lambda$ and use
the aforementioned convexity diagnostics to determine the locally
convex regions of the solution path.  If the chosen solution lies in
the region below $\lambda^*$, increase $\gamma$ to make the penalty
more convex.  On the other hand, if the chosen solution lies well above
$\lambda^*$, one can lower $\gamma$ without fear of nonconvexity.  Once
this process has been iterated a few times to find a value of $\gamma$
that seems to produce an appropriate balance between parsimony and
convexity, use cross-validation to choose $\lambda$ for this value of
$\gamma$.  This approach is illustrated in Section~\ref{section:app}.

%s5 ###
\section{Numerical results}
\label{section:sim}

%s5.1 ###
\subsection{Computational efficiency}

In this section we assess the computational efficiency of the
coordinate descent and LLA algorithms for fitting MCP and SCAD
regression models.  We examine the time required to fit the entire
coefficient path for linear and logistic regression models.

In these simulations, the response for linear regression was generated
from the standard normal distribution, while for logistic regression,
the response was generated as a Bernoulli random variable with
$P(y_i=1)=0.5$ for all $i$.

To investigate whether or not the coordinate descent algorithm
experiences difficulty in the presence of correlated predictors,
covariate values were generated in one of two ways: independently from
the standard normal distribution (i.e., no correlation between
the covariates), and from a multivariate normal distribution with a
pairwise correlation of $\rho=0.9$ between all covariates.

\begin{figure}

\includegraphics{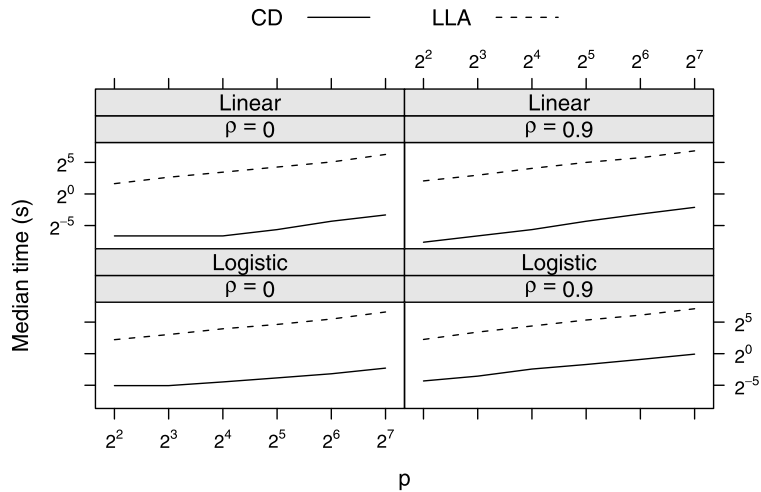}

\caption{Time required (in seconds) to fit the entire
coefficient paths for linear and logistic regression, employing either
the local linear approximation (LLA) algorithm or the coordinate
descent (CD) algorithm.  Both axes are on the log scale.  Times
displayed are averaged over 100 independently generated data sets.  The
coordinate descent algorithm is at least 100 times faster than the LLA
algorithm at all points.}\label{fig:eff}
\end{figure}

To ensure that the comparison between the algorithms was fair and not
unduly influenced by failures to converge brought on by nonconvexity or
model saturation, $n$ was chosen to equal 1000 and $\gamma$ was set
equal to $1/c_*$, thereby ensuring a convex objective function for
linear regression.  This does not necessarily ensure that the logistic
regression objective function is convex, although with adaptive
rescaling, it works reasonably well.  In all of the cases investigated,
the LLA and coordinate descent algorithms converged to the same path
(within the accuracy of the convergence criteria).  The median times
required to fit the entire coefficient path are presented as a function
of $p$ in Figure~\ref{fig:eff}.

%f3 ###

Interestingly, the slope of both lines in Figure~\ref{fig:eff} is close
to 1 in each setting (on the log scale), implying that both coordinate
descent and LLA exhibit a linear increase in computational burden with
respect to $p$ over the range of $p$ investigated here.  However,
coordinate descent algorithm is drastically more efficient---up to
1000 times faster.  The coordinate descent algorithm is somewhat (2--5
times) slower in the presence of highly correlated covariates, although
it is still at least 100 times faster than LLA in all settings
investigated.

%s5.2 ###
\subsection{Comparison of MCP, SCAD and lasso}

We now turn our attention to the statistical properties of MCP and SCAD
in comparison with the lasso.  We will examine two instructive sets of
simulations, comparing these nonconvex penalty methods to the lasso.
The first set is a simple series designed to illustrate the primary
differences between the methods.  The second set is more complex, and
designed to mimic the applications to real data in
Section~\ref{section:app}.

In the simple settings, covariate values were generated independently
from the standard normal distribution.  The sample sizes were $n=100$
for linear regression and $n=400$ for logistic regression.  In the
complex settings, the design matrices from the actual data sets were
used.

For each data set, the MCP and SCAD coefficient paths were fit using
the coordinate descent algorithm described in this paper, and lasso
paths were fit using the \texttt{glmnet} package [\citet{Friedman2010}].
Tenfold cross-validation was used to choose $\lambda$.

\textit{Setting }1.  We begin with a collection of simple models in which
there are four nonzero coefficients, two of which equal to $+s$, the
other two equal to $-s$.  Given these values of the coefficient vector,
responses were generated from the normal distribution with mean
$\eta_i$ and variance 1 for linear regression, and for logistic
regression, responses were generated according to \eqref{eq:log-model}.

For the simulations in this setting, we used the single value
$\gamma=3$ in order to illustrate that reasonable results can be
obtained without investigation of the tuning parameter if adaptive
rescaling is used for penalized logistic regression.  Presumably, the
performance of both MCP and SCAD would be improved if $\gamma$ was
chosen in a more careful manner, tailored to each setting;
nevertheless, the simulations succeed in demonstrating the primary
statistical differences between the methods.

\begin{figure}[b]

\includegraphics{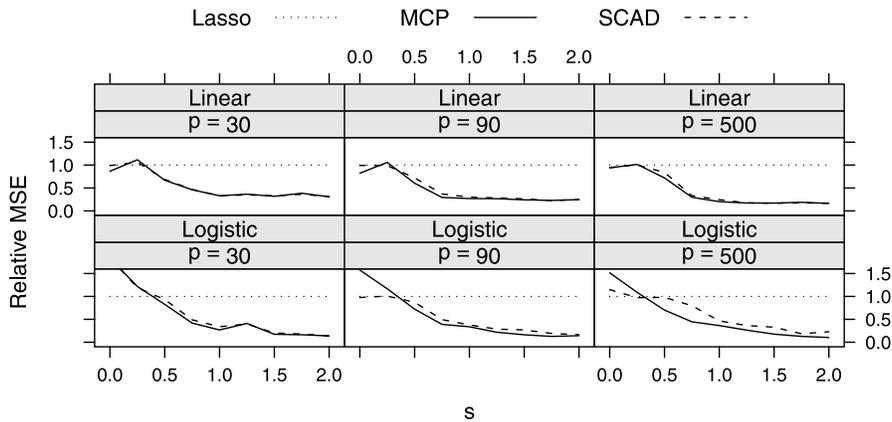}

\caption{Relative (to the lasso) mean squared error
for MCP- and SCAD-penalized linear and logistic regression.  MSE was
calculated for each penalty on 100 independently generated data sets,
and the ratio of the medians is plotted.  MCP and SCAD greatly
outperform lasso for large coefficients, but not necessarily for small
coefficients.}\label{fig:cmp1}
\end{figure}

We investigate the estimation efficiency of MCP, SCAD and lasso as $s$
and $p$ are varied.  These results are shown in Figure~\ref{fig:cmp1}.

%f4 ###

Figure~\ref{fig:cmp1} illustrates the primary difference between
MCP/SCAD and lasso.  MCP and SCAD are more aggressive modeling
strategies, in the sense that they allow coefficients to take on large
values much more easily than lasso.  As a result, they are capable of
greatly outperforming the lasso when large coefficients are present in
the underlying model.  However, the shrinkage applied by the lasso is
beneficial when coefficients are small, as seen in the regions of
Figure~\ref{fig:cmp1} in which $s$ is near zero.  In such settings, MCP
and SCAD are more likely to overfit the noisy data.  This is true for
both linear and logistic regression.  This tendency is worse for
logistic regression than it is for linear regression, worse for MCP
than it is for SCAD, and worse when $\gamma$ is closer to 1 (results
supporting this final comparison not shown).

\textit{Setting }2.  In this setting we examine the performance of MCP,
SCAD and lasso for the kinds of design matrices often seen in
biomedical applications, where covariates contain complicated patterns
of correlation and high levels of multicollinearity.
Section~\ref{section:app} describes in greater detail the two data
sets, which we will refer to here as ``Genetic Association'' and
``Microarray,'' but it is worth noting here that for the Genetic
Association case, $n=800$ and $p=532$, while for the Microarray case,
$n=38$ and $p=7129$.

In both simulations the design matrix was held constant while the
coefficient and response vectors changed from replication to
replication.  In the Genetic Association case, 5 coefficient values
were randomly generated from the exponential distribution with rate
parameter 3, given a random ($+/-$) sign, and randomly assigned to one
of the 532 SNPs (the rest of the coefficients were zero).  In the
Microarray case, 100 coefficient values were randomly generated from
the normal distribution with mean 0 and standard deviation 3, and
randomly assigned among the 7129 features.  Once the coefficient
vectors were generated, responses were generated according to
\eqref{eq:log-model}.  This was repeated 500 times for each case, and
the results are displayed in Table~\ref{tab:appsim}.

%t1 ###
\begin{table}[b]
\tablewidth=260pt
\caption{Simulation results: Setting \textup{2}\% correct
refers to the percent of variables selected by the model that had
nonzero coefficients in the generating model}\label{tab:appsim}
\begin{tabular}{@{}lccc@{}}
\hline
&\textbf{Misclassification}&&\\
\textbf{Penalty}&\textbf{error (\%)}&\textbf{Model size}&\textbf{Correct (\%)}\\
\hline
\textit{Genetic association}&&&\\
\quad MCP ($\gamma=1.5$)&38.7&\phantom{0}2.1&54.7\\
\quad SCAD ($\gamma=2.5$)&38.7&\phantom{0}9.5&25.0\\
\quad Lasso&38.8&12.6&20.6\\[3pt]
\textit{Microarray}&&&\\
\quad MCP ($\gamma=5$)&19.7&\phantom{0}4.3&\phantom{0}3.2\\
\quad MCP ($\gamma=20$)&16.7&\phantom{0}7.5&\phantom{0}4.3\\
\quad SCAD ($\gamma=20$)&16.4&\phantom{0}8.7&\phantom{0}4.3\\
\quad Lasso&16.2&\phantom{0}8.8&\phantom{0}4.3\\
\hline
\end{tabular}
\end{table}

The Genetic Association simulation---reflecting the presumed
underlying biology in these kinds of studies---was designed to be
quite sparse.  We expected the more sparse MCP method to outperform the
lasso here, and it does.  All methods achieve similar predictive
accuracy, but MCP does so using far fewer variables and with a much
lower percentage of spurious covariates in the model.

The Microarray simulation was designed to be unfavorable to sparse
model\-ing---a large number of nonzero coefficients, many of which are
close to zero.  As is further discussed in Section~\ref{section:app},
$\gamma$ should be large in this case; when $\gamma=5$, MCP produces
models with diminished accuracy in terms of both prediction and
variable selection.  However, it is not the case that lasso is clearly
superior to MCP here; if $\gamma$ is chosen appropriately, the
predictive accuracy of the lasso can be retained, with gains in
parsimony, by applying MCP with $\gamma=20$.

In the microarray simulation, the percentage of variables selected that
have a nonzero coefficient in the generating model is quite low for all
methods.  This is largely due to high correlation among the features.
The methods achieve good predictive accuracy by finding predictors that
are highly correlated with true covariates, but are unable to
distinguish the causative factors from the rest in the network of
interdependent gene expression levels.  This is not an impediment for
screening and diagnostic applications, although it may or may not be
helpful for elucidating the pathway of disease.

SCAD, the design of which has similarities to both MCP and lasso, lies
in between the two methodologies.  However, in these two cases, SCAD is
more similar to lasso than it is to MCP.

%s6 ###
\section{Applications}
\label{section:app}

%s6.1 ###
\subsection{Genetic association}

Genetic association studies have become a wide\-ly used tool for
detecting links between genetic markers and diseases.  The example that
we will consider here involves data from a case-control study of
age-related macular degeneration consisting of 400 cases and 400
controls.  We confine our analysis to 30 genes that previous biological
studies have suggested may be related to the disease.  These genes
contained 532 markers with acceptably high call rates and minor allele
frequencies.  Logistic regression models penalized by lasso, SCAD and
MCP were fit to the data assuming an additive effect for all markers
(i.e., for a homozygous dominant marker, $x_{ij}=2$, for a
heterozygous marker, $x_{ij}=1$, and for a homozygous recessive marker,
$x_{ij}=0$).  As an illustration of computational savings in a
practical setting, the LLA algorithm required 17 minutes to produce a
single MCP path of solutions, as compared to 2.5 seconds for coordinate
descent.

\begin{figure}

\includegraphics{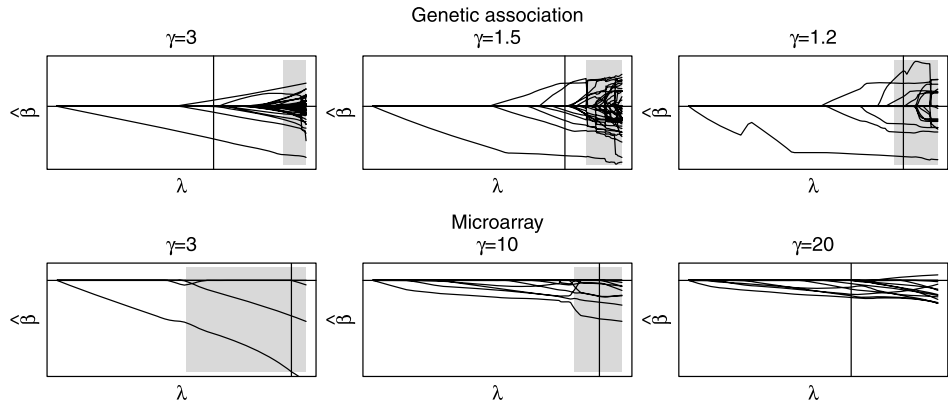}

\caption{MCP coefficient paths
for various values of $\gamma$ for the two studies of
Section~\protect\ref{section:app}.  The shaded region depicts areas that are
not locally convex, and a vertical line is drawn at the value of
$\lambda$ selected by BIC.  For the sparse genetic association data,
small values of $\gamma$ produce the best fit; for the dense microarray
data, large values are preferred.}\label{fig:app-gamma}
\end{figure}

As described in Section~\ref{section:selection}, we used BIC and
convexity diagnostics to choose an appropriate value of $\gamma$; this
is illustrated in the top half of Figure~\ref{fig:app-gamma}.  The
biological expectation is that few markers are associated with the
disease; this is observed empirically as well, as a low value of
$\gamma=1.5$ was observed to balance sparsity and convexity in this
setting.  In a similar fashion, $\gamma=2.5$ was chosen for SCAD.

%f5 ###

Ten-fold cross-validation was then used to choose $\lambda$ for MCP,
SCAD and lasso.  The number of parameters in the selected model, along
with the corresponding cross-validation errors, are listed in
Table~\ref{tab:ga}.  The results indicate that MCP is better suited to
handle this sparse regression problem than either SCAD or lasso,
achieving a modest reduction in cross-validated prediction error while
producing a much more sparse model.  As one would expect, the SCAD
results are in between MCP and lasso.

%t2 ###
\begin{table}[b]
 \caption{Genetic association results}\label{tab:ga}
\begin{tabular}{@{}lcc@{}}
\hline
\textbf{Penalty}&\textbf{Model size}&\textbf{CV error (\%)}\\
\hline
MCP&\phantom{00}7&39.4\\
SCAD&\phantom{0}25&40.6\\
Lasso&103&40.9\\
\hline
\end{tabular}
\end{table}

%s6.2 ###
\subsection{Gene expression}

Next, we examine the gene expression study of leuke\-mia patients
presented in \citet{Golub1999}.  In the study the expression levels of
7129 genes were recorded for 27 patients with acute lymphoblastic
leukemia (ALL) and 11 patients with acute myeloid leukemia (AML).
Expression levels for an additional 34 patients were measured and
reserved as a test set.  Logistic regression models penalized by lasso,
SCAD and MCP were fit to the training data.

Biologically, this problem is more dense than the earlier application.
Potentially, a large number of genes are affected by the two types of
leukemia.  In addition, the sample size is much smaller for this
problem.  These two factors suggest that a higher value of $\gamma$ is
appropriate, an intuition borne out in Figure~\ref{fig:app-gamma}.  The
figure suggests that $\gamma \approx 20$ may be needed in order to
obtain an adequately convex objective function for this problem (we
used $\gamma=20$ for both MCP and SCAD).

With $\gamma$ so large, MCP and SCAD are quite similar to lasso;
indeed, all three methods classify the test set observations in the
same way, correctly classifying 31$/$34.  Analyzing the same data,
\citet{Park2007} find that lasso-penalized logistic regression is
comparable with or more accurate than several other competing methods
often applied to high-dimensional classification problems.  The same
would therefore apply to MCP and SCAD as well; however, MCP achieved
its success using only 11 predictors, compared to 13 for lasso and
SCAD.  This is an important consideration for screening and diagnostic
applications such as this one, where the goal is often to develop an
accurate test using as few features as possible in order to control
cost.

Note that, even in a problem such as this one with a sample size of 38
and a dozen features selected, there may still be an advantage to the
sparsity of MCP and the parsimonious models it produces.  To take
advantage of MCP, however, it is essential to choose $\gamma$ wisely---using the value $\gamma=5$ (much too sparse for this problem) tripled
the test error to 9$/$34.  We are not aware of any method that can
achieve prediction accuracy comparable to MCP while using only 11
features or fewer.

%s7 ###
\section{Discussion}

The results from the simulation studies and data examples considered in
this paper provide compelling evidence that nonconvex penalties like
MCP and SCAD are worthwhile alternatives to the lasso in many
applications.  In particular, the numerical results suggest that MCP is
often the preferred approach among the three methods.

Many researchers and practitioners have been reluctant to embrace these
methods due to their lack of convexity, and for good reason: nonconvex
objective functions are difficult to optimize and often produce
unstable solutions.  However, we provide here a fast, efficient and
stable algorithm for fitting MCP and SCAD models, as well as introduce
diagnostic measures to indicate which regions of a coefficient path are
locally convex and which are not.  Furthermore, we introduce an
adaptive rescaling for logistic regression which makes selection of the
tuning parameter $\gamma$ much easier and more intuitive.  All of these
innovations are publicly available as an open-source \texttt{R} package
(\url{http://cran.r-project.org}) called \texttt{ncvreg}.  We hope
that these efforts remove some of the barriers to the further study and
use of these methods in both research and practice.

\begin{appendix}\label{app}
\section*{Appendix}

Although the objective functions under consideration in this paper are
not differentiable, they possess directional derivatives and
directional second derivatives at all points $\bb$ and in all
directions $\u$ for $\bb, \u \in \mathbb{R}^p$.  We use $d_{\u}Q$ and
$d_{\u}^2Q$ to represent the derivative and second derivative of $Q$ in
the direction $\u$.

\begin{pf*}{Proof of Lemma \ref{1lema}}
For all $\beta_j \in (-\infty,\infty)$,
\[
\min\{d_{-}^2Q_{j,\lambda,\gamma}(\beta),d_{+}^2Q_{j,\lambda,\gamma}(\beta)\} \geq
\cases{
\displaystyle 1-\frac{1}{\gamma}&\quad for MCP,\cr
\displaystyle 1-\frac{1}{\gamma-1}&\quad for SCAD.}
\]
Thus, $Q_{j,\lambda,\gamma}(\beta)$ is a strictly convex function on
$(-\infty,\infty)$ if $\gamma>1$ for MCP and if $\gamma>2$ for SCAD.
\end{pf*}

\begin{pf*}{Proof of Proposition \ref{prop:cd}}
\citet{Tseng2001} establishes
sufficient conditions for the convergence of cyclic coordinate descent
algorithms to coordinate-wise minima.  The strict convexity in each
coordinate direction established in Lemma 1 satisfies the conditions of
Theorems 4.1 and 5.1 of that article.  Because $Q$ is continuous,
either theorem can be directly applied to demonstrate that the
coordinate descent algorithms proposed in this paper converge to
coordinate-wise minima.  Furthermore, because all directional
derivatives exist, every coordinate-wise minimum is also a local
minimum.
\end{pf*}

\begin{pf*}{Proof of Proposition \ref{prop:logistic-convex}}
For all $\bb \in \mathbb{R}^p$,
\[
\min\limits_{\u}\{d_{\u}^2Q_{\lambda,\gamma}(\bb)\} \geq
\cases{
\displaystyle\frac{1}{n} \X'\W\X - \frac{1}{\gamma}\I&\quad for MCP,\cr
\displaystyle\frac{1}{n} \X'\W\X - \frac{1}{\gamma-1}\I&\quad for SCAD.}
\]
Thus, $Q_{\lambda,\gamma}(\bb)$ is a convex function on the region for
which $c_*(\bb) > 1/\gamma$ for MCP, and where $c_*(\bb) >
1/(\gamma-1)$ for SCAD.
\end{pf*}
\end{appendix}

\section*{Acknowledgments}

The authors would like to thank Professor Cun-Hui Zhang for providing
us an early version of his paper on MCP and sharing his insights on the
related topics, and Rob Mullins for the genetic association data
analyzed in Section 6.  We also thank the editor, associate editor and
two referees for comments that improved both the clarity of the writing
and the quality of the research.

\printaddresses

\end{document}